# Screw dislocation interaction in smectic A liquid crystals in an anharmonic approximation

Lubor LEJČEK

Institute of Physics, Czech Academy of Sciences, Prague, Czech Republic

(lejcekl@fzu.cz)

ABSTRACT

The interaction of two screw dislocations in smectic-A liquid crystals is treated using an anharmonic correction to the elastic energy density. In the present contribution the elastic energy of two screw dislocations is evaluated and discussed. For screw dislocations with both parallel and opposite Burgers vectors there is an attraction of dislocations for their small separations while for greater separations there is a repulsion. It can be explained by dominated terms in interaction energy which do not depend on signs of dislocations. In this way, the interaction energy of screw dislocations in smectic A liquid crystal within an anharmonic approximation differs with respect to the case of screw dislocations in solids.

1. ## Introduction

The classical solution describing a screw dislocation in the system of smectic layers of the smectic A liquid crystal is known for long time [1 - 4]. This classical solution determined within the smectic elasticity described by the elastic constant of layer curvature $K$ and the layer compression $B$ in infinite medium has the zero self-energy. Similarly, the interaction energy of two screw dislocations calculated in infinite medium is also zero [3,4].

On the other hand, when an anharmonic correction of layer compression to the smectic elasticity is taken in account, the contribution to the elastic self-energy of screw dislocation is non-zero [3, 4, 6]. This anharmonic contribution was also used to determine the critical undulations of smectic layers subjected to the dilatational deformations [4].

Screw dislocations were treated not only in linear approach as in [1,2] but also in non-linear approach, see e.g. [5, 7, 8]. Then the screw dislocation interaction in non-linear approach can be approximately determined [5] giving approximately logarithmic dependence for dislocation separations much greater then dislocation core radius.

In [7] the energy of a wall formed of parallel screw dislocations in the smectic-A was evaluated using anharmonic term and discussed in the connection with twist-grain-boundaries.

A classical solution of screw dislocation was reexamined in [9]. There screw dislocation deformation is modelled in a hollow cylinder of inner radius $r_c$ and outer radius $R$ with proper boundary conditions. In [9] a solution describing the screw dislocation in the cylinder of finite radius $R$ contains other terms with respect to the classical solution [1 - 4]. It should be noted, however, that in the limit of infinite outer radius $R$, i.e. $R \to \infty$, the solution given in [9] is transformed to the classical solution for the screw dislocation. When calculating the screw dislocation self-energy, Ref. [9] considers not only the classical elastic energy density but also adds the term proportional to the anharmonic correction to the energy (see expression $U_2$ in (17) of [9]). For this reason the energy $U_2$ of [9] leads to the Kléman´s contribution [6] of the screw dislocation energy in the limit $R \to \infty$.

In the present contribution we will examine the classical problem of the interaction of two isolated screw dislocations in infinite smectic A liquid crystal. The anharmonic approximation [4] used in this contribution gives the expression for the interaction energy and for the force between dislocations in the analytical form. Anyway, the anharmonic term is just a corrective term. It gives interaction forces between screw dislocations about of two orders smaller as compared with edge dislocation interaction forces.

In this contribution the screw dislocation core is not treated as it was well discussed e.g. in [10, 11].

## 2. The free energy density of smectic A liquid crystal

Let the system of smectic layers is parallel to the plane $(x, y)$ and the layer normal oriented along the $z$-axis is perpendicular to layers (Fig. 1). The free energy density $f$ of smectic A liquid crystal can be written in the well-known form [3, 4, 12]:

$$f = \frac{B}{2}\left(\frac{\partial u}{\partial z}\right)^2 + \frac{K}{2}\left(\frac{\partial^2 u}{\partial x^2} + \frac{\partial^2 u}{\partial y^2}\right)^2, \qquad (1)$$

where $u$ is the layer displacement in $z$-direction. Parameter $B$ is the layer compression elastic constant and $K$ is the elastic constant characterizing the smectic layer curvature in one-constant approximation.

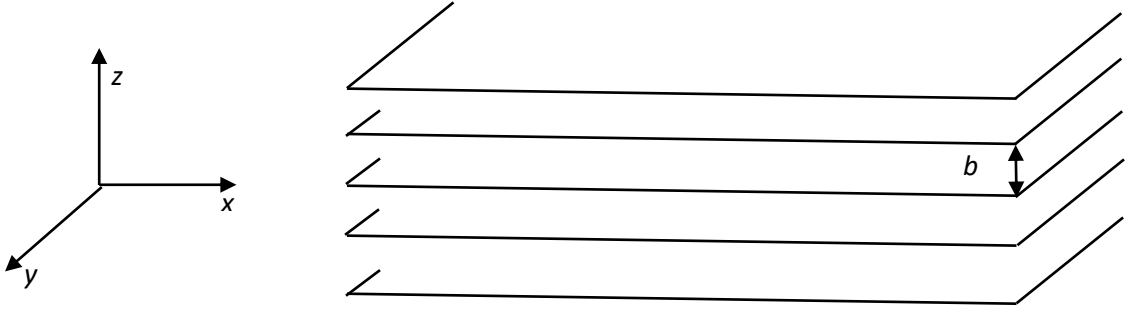

Fig. 1: Smectic A layers and coordinate system. $b$ is the smectic layer thickness and the Burgers vector.

Supposing the screw dislocation be parallel to the $z$-direction we expect that the dislocation displacement does not depend on the $z$ coordinate. Therefore the calculus of variation applied to the dislocation energy

$$\int_V f \, dV, \tag{2}$$

leads to the equilibrium equation:

$$\Delta\Delta u = 0, \tag{3}$$

where $\Delta = \frac{\partial^2}{\partial x^2} + \frac{\partial^2}{\partial y^2}$ and $u = u(x, y)$.

The classical solution describing the screw dislocation in an infinite smectic A liquid crystal was fond in form [1 - 4]:

$$u = \frac{b}{2\pi} Arctan \frac{y}{x}, \tag{4}$$

with $b$ as the value of Burgers vector oriented along the $z$-direction. The solution (4) satisfies the condition:

$$\oint_\Gamma du = b, \tag{5}$$

where the integral taken along the closed loop $\Gamma$ around the dislocation core gives the value of Burgers vector $b$.

Eq. (1) immediately shows that the self-energy of the screw dislocation (4) is zero. Therefore, to find a contribution to the non-zero self-energy and thus the contribution to the elastic interaction energy of two screw dislocation, we add the anharmonic term in the form [4]:

$$\frac{B}{8} \iint \left( \left(\frac{\partial u}{\partial x}\right)^2 + \left(\frac{\partial u}{\partial y}\right)^2 \right)^2 dxdy. \tag{6}$$

In (6) we integrate over the surface of smectic layer (note that energy (6) is the energy per unit length of dislocation in z-direction).

The insertion of solution (4) into (6) and the use of polar coordinates $x = r\cos\varphi$, $y = r\sin\varphi$ permits the evaluation of integral (6) as:

$$\frac{Bb^4}{128\pi^4} \int_0^{2\pi} d\varphi \int_{r_c}^{R} \frac{rdr}{r^4} = \frac{Bb^4}{128\pi^3}\left(\frac{1}{r_c^2} - \frac{1}{R^2}\right) \to \frac{Bb^4}{128\pi^3 r_c^2}, \tag{7}$$

in the limit $R \to \infty$. Parameter $r_c$ is the core radius introduced to avoid the singularity at $r = 0$. The energy given by (7) is the contribution of the screw dislocation self-energy in anharmonic approximation first introduced by Kléman [6] (see also [4]).

### 3. Elastic energy of two screw dislocations in anharmonic approximation

In this contribution, we use the classical solution (4) for description of screw dislocations valid for an infinite sample. It is the sufficient approximation of the solution because the solution of [9] gives (4) in the limit $R \to \infty$.

Now, the integral (6) will be evaluated for two screw dislocations, one situated at the coordinate origin and described by the solution $u_1 = \frac{b_1}{2\pi} Arctan \frac{y}{x}$ with Burgers vector $b_1$ and the other by the solution $u_2 = \frac{b_2}{2\pi} Arctan \frac{y-y_o}{x-x_o}$ with Burgers vector $b_2$ situated at coordinates $x_o$ and $y_o$. The coordinates $x_o$ and $y_o$ can be written in polar coordinates as: $x_o = r_o \cos\varphi_o$ and $y_o = r_o \sin\varphi_o$. The radial distance between two dislocations in polar coordinates is $r_o$.

Then the solution describing two screw dislocations can be written as:

$$u = u_1 + u_2. \tag{8}$$

The anharmonic contribution $\frac{B}{8}\left(\left(\frac{\partial u}{\partial x}\right)^2 + \left(\frac{\partial u}{\partial y}\right)^2\right)^2$ with (8) and in polar coordinates is expressed as:

$$\frac{B}{128\pi^4} \frac{\left(2b_1 b_2 r^2 + b_2^2 r^2 + b_1^2(r^2+r_o^2) - 2b_1(b_1+b_2)rr_o\cos(\varphi-\varphi_o)\right)^2}{r^4\left(r^2+r_o^2-2rr_o\cos^2(\varphi-\varphi_o)\right)^2}. \tag{9}$$

First, we integrate expression (9) by the coordinate $\varphi$ in the interval $\varphi \in (0,2\pi)$. As the screw dislocation with the Burgers vector $b_2$ is situated at $\varphi_o$, we avoid the integration over the interval $(\varphi_o - \varphi_c, \varphi_o + \varphi_c)$ where $\varphi_c$ is a small cut-off angle. The integration using Mathematica [13] shows that the evaluation of integral leads to the term of the type $Arctan\left(\left(\frac{r+r_o}{r-r_o}\right) Tan \frac{\varphi-\varphi_o}{2}\right)$. This expression near point $\pi - \varphi_o$ should be evaluated as the difference of limits approaching from the smaller values of $\varphi$ and from larger values of $\varphi$. The ratio $\left(\frac{r+r_o}{r-r_o}\right)$ leads in the evaluation to the function $Sgn\left(\frac{r+r_o}{r-r_o}\right)$. Finally, Mathematica gives the integral over $\varphi \in (0,2\pi)$ in the form:

$$I_{r_o} = \frac{B}{128\pi^4 r^4} \left\{ -\frac{2(b_2 r + b_1(r - r_o))^4}{(r - r_o)^4} \varphi_c + 2\pi b_1^2 (b_1 + b_2)^2 + \right.$$

$$\left. \pi Sgn\left(\frac{r+r_o}{r-r_o}\right) \left( 4b_1^3 b_2 + 8b_1 b_2^3 \frac{r^4}{(r^2-r_o^2)^2} + 2b_1^2 b_2^2 \frac{(5r^2-r_o^2)}{(r^2-r_o^2)} + 2b_2^4 \frac{r^4(r^2+r_o^2)}{(r^2-r_o^2)^3} \right) \right\}. \quad 10)$$

In evaluation of (10) we developed the obtained expression up to the first order of $\varphi_c$ as we expect $\varphi_c$ to be a small parameter.

Expression (10) has the singularities at $r = 0$ and at $r = r_o$ where dislocations are situated. The total elastic energy of two screw dislocations, $E_T$, can be evaluated as the integral:

$$E_T = \int_{r_c}^{r_o - r_c} I_{r_o} r \, dr + \int_{r_o + r_c}^{\infty} I_{r_o} r \, dr. \quad (11)$$

Expression (11) is the dislocation elastic self-energy per unit length of dislocation in $z$-direction. Parameter $r_c$ in expression (11) is used to avoid the singularities. Integral (11) is divided into two parts as $Sgn\left(\frac{r+r_o}{r-r_o}\right) = -1$ in the interval $r \in (r_c, r_o - r_c)$ and $Sgn\left(\frac{r+r_o}{r-r_o}\right) = +1$ in the interval $r \in (r_o + r_c, \infty)$.

Using (11) we can test limits of $E_T$ for small separation $r_o$ of dislocations and for greater separation $r_o \to \infty$. The first integral in (11) is zero when $r_o \to 2r_c$. The second integral in (11) does not diverge for $r_o \to 0$ and gives $\frac{B(b_1+b_2)^4}{128\pi^3 r_c^2}$ which is the self-energy of two screw dislocations with Burgers vectors $b_1$ and $b_2$ situated at the coordinate origin. On the other hand, for $r_o \to \infty$ the second integral in (11) is zero, while the first integral gives $\frac{B b_1^4}{128\pi^3 r_c^2}$. The last expression is the self-energy of the screw dislocation with the Burgers vector $b_1$ at the coordinate origin. The dislocation with $b_2$ does not contribute as it is displaced to infinity.

Before integration, let us substitute formally $\varphi_c$ to $\varphi_c \to p_r/r_o$. The parameter $p_r$ will be determined later.

Using Mathematica [13], integral (11) can be evaluated as:

$$E_T = \frac{B}{128\pi^4} \left\{ -\frac{4b_2^4 p_r}{3r_c^3} + \frac{b_1^4}{r_c^2}\left(\pi - \frac{p_r}{r_o}\right) + \frac{b_2^3(4b_1+b_2)\pi}{r_c(r_c-2r_o)} - \frac{2b_2^4 p_r}{3(r_c-r_o)^3} \right.$$

$$+ \frac{8b_1^3 b_2 p_r - 24 b_1^2 b_2^2 p_r}{r_c r_o^2} + \frac{8b_1^3 b_2 p_r - 12 b_1^2 b_2^2 p_r}{(r_c - r_o) r_o^2} - \frac{b_2^3(4b_1+b_2) p_r}{(r_c - r_o)^2 r_o} + \frac{b_2^4 \pi r_o^2}{r_c^2(r_c - 2r_o)^2}$$

$$+ \frac{b_1^2 \left(4b_1 b_2 \pi + 2 b_2^2 \pi + b_1^2 \left(\pi - \frac{p_r}{r_o}\right)\right)}{(r_c + r_o)^2} + \frac{8b_1^3 b_2 p_r}{(r_c + r_o) r_o^2} + \frac{b_2^4 \pi r_o^2}{r_c^2(r_c + 2r_o)^2}$$

$$\left. + \frac{b_2^3(4b_1 + b_2)\pi}{r_c(r_c + 2r_o)} \right.$$

$$+\frac{b_1^4}{(r_c-r_o)^2 r_o}(p_r-\pi r_o)-\frac{b_2^4 \pi r_0^2}{(r_c^2-r_0^2)^2}-\frac{b_2^3(4b_1+b_2)\pi}{(r_c^2-r_0^2)}$$

$$+\frac{4b_1^2 b_2}{r_o^3}\left(-b_2\pi r_o \ln\left|1-\frac{2r_o}{r_c}\right|+2(2b_1 p_r-3b_2 p_r+b_2\pi r_o)\ln\left|\frac{r_o}{r_c}-1\right|+\right.$$

$$\left. b_2\pi r_o \ln\left|\left(\frac{r_o}{r_c}\right)^2-1\right|\right)\bigg\}. \quad (12)$$

Note that energy $E_T$ is the energy per unit length of dislocation.

Using (12) we can now determine the formally introduced parameter $p_r$. Limit of (12) for $r_o \to \infty$ gives $\frac{B}{1536\pi^4 r_c^3}(12b_1^4\pi r_c + b_2^4(-8p_r+3\pi r_c))$. This expression will be $\frac{Bb_1^4}{128\pi^3 r_c^2}$ if we put $p_r = 3\pi r_c/8$. It also means that the angle $\varphi_c$ can be expressed as $\varphi_c = 3\pi r_c/8r_o$. So $\varphi_c$ will be small value for higher $r_o$.

Using parameters $p_r = 3\pi r_c/8$, $t = \frac{r_o}{r_c}$, $b_1 = ba_1$, $b_2 = ba_2$, where $b$ is the absolute value of Burgers vector and $a_1$ and $a_2$ are $\pm 1$ we substitute them into (12) and we obtain:

$$\frac{E_T}{\left(\frac{Bb^4}{1024\pi^3 r_c^2}\right)} = -4a_2^4 + \frac{2a_2^4}{(t-1)^3} + \frac{24a_1^3 a_2}{t^2} - \frac{72a_1^2 a_2^2}{t^2} - \frac{24a_1^3 a_2}{t^2(t-1)} + \frac{36a_1^2 a_2^2}{t^2(t-1)} -$$

$$\frac{3a_2^3(4a_1+a_2)}{(t-1)^2 t} + \frac{8a_2^4 t^2}{(1-2t)^2} + \frac{24a_1^3 a_2}{t^2(t+1)} - \frac{8a_2^3(4a_1+a_2)}{(2t-1)} + \frac{8a_2^4 t^2}{(1+2t)^2}$$

$$+ \frac{8a_2^3(4a_1+a_2)}{(2t+1)} +$$

$$a_1^4(8t-3)\left(\frac{1}{t}-\frac{1}{(t-1)^2 t}\right) - \frac{8a_2^4 t^2}{(-1+t^2)^2} + \frac{8a_2^3(4a_1+a_2)}{(-1+t^2)} +$$

$$\frac{a_1^2(32a_1 a_2 t + 16a_2^2 t + a_1^2(8t-3))}{(t+1)^2 t} +$$

$$\frac{4a_1^2 a_2}{t^3}(-8a_2 t\ln|2t+1| - 8a_2 t\ln|2t-1| + (12a_1-18a_2+16a_2 t)\ln|t-1| +$$

$$(-6a_1+9a_2+16a_2 t)\ln|t+1| + 8a_2 t\ln|t^2-1|). \quad (13)$$

Using the expression (13) the elastic energy of two parallel screw dislocation can be displayed as a function of the parameter $t = \frac{r_o}{r_c}$. To avoid a singularity, $t \geq 2$ because of integration limits in the first integral of (11). For dislocations with the parallel Burgers vectors, i.e. for $a_1 = a_2 = 1$ the elastic energy is shown in Fig. 2:

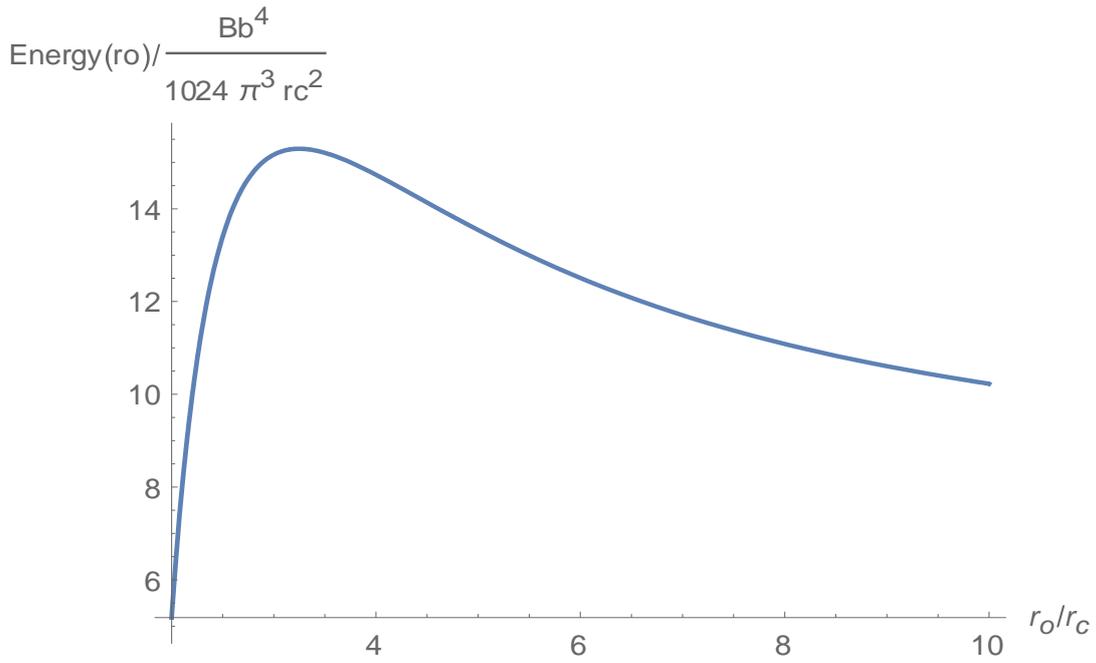

Fig. 2: The elastic energy of two screw dislocations of the same sign as the function of their separation $\frac{r_o}{r_c}$.

For dislocations with the antiparallel parallel Burgers vectors, i.e. for $a_1 = 1$ and $a_2 = -1$ or $a_1 = -1$ and $a_2 = 1$ the elastic energy is shown in Fig. 3:

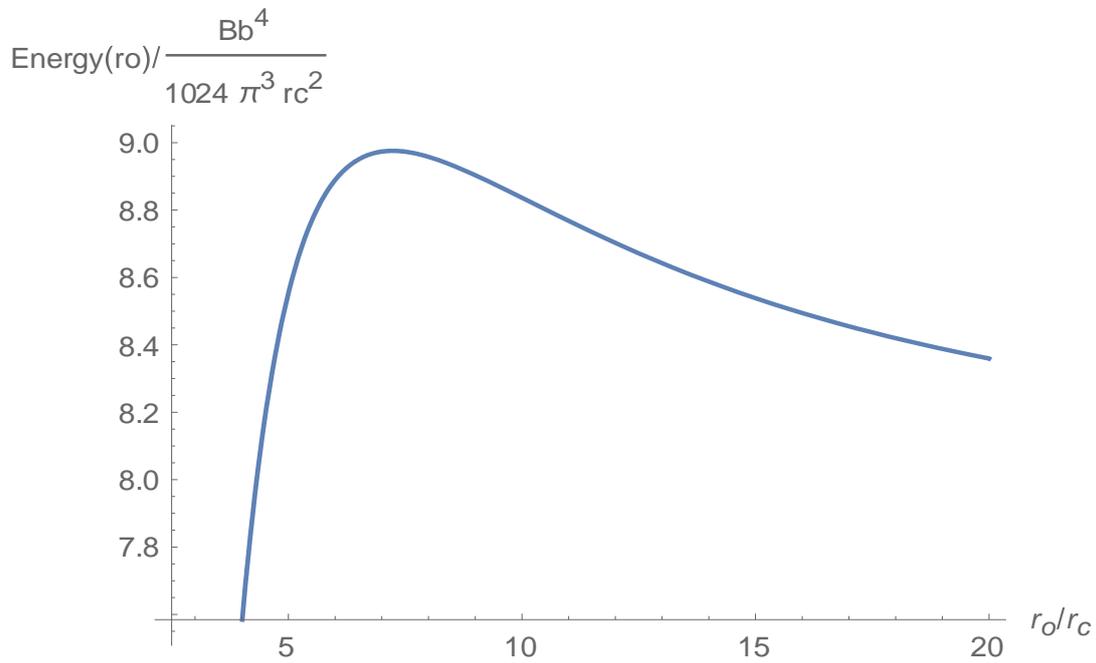

Fig. 3: The elastic energy of two screw dislocations of the opposite sign as the function of their separation $\frac{r_o}{r_c}$.

## 4. Discussion

Total elastic energy of two screw dislocations was calculated using the nonlinear anharmonic term. As seen from Figs. 2 and 3, for both parallel and antiparallel screw dislocations there is a maximum for some non-zero separation of dislocations. It means that the interaction of screw dislocations in the anharmonic approximation differs from the behavior of screw dislocations e.g. in solids, described by linear elasticity. In solids, the dislocation interaction energy is usually monotonic function while in anharmonic approximation of smectic A we observe a maximum of the energy which does not depends importantly on the dislocation signs. In the energy expression (13) the interaction terms which do not change with dislocation signs are proportional to $b_1^2 b_2^2$ (or $b^2 a_1^2 a_2^2$). These terms are the consequence of non-linearity of the anharmonic term and they lead to energy maximum. Therefore they can influence the character of dislocation interaction as was already expected and predicted by Pleiner [5].

Due to the existence of this energy maximum, in both cases of screw dislocations with either parallel or antiparallel Burgers vectors there is a dislocation attraction for small dislocation separation ($\frac{r_o}{r_c} < 3{,}25$ for parallel Burgers vectors and $\frac{r_o}{r_c} < 7{,}24$ for antiparallel Burgers vectors). The attraction is unexpected for parallel Burgers vectors as it could lead to possible coalescence of dislocations. As remarked in [4], most screw dislocations are elementary. It could be also understood by the fact that they usually cannot approach to distances lower then about the $3 r_c$ due to the energy barrier. For greater distances there is the dislocation repulsion what can be expected.

As for screw dislocations of opposite Burgers vectors the expected attraction exists just at their very short separations. For greater separations there is a small repulsion (see Fig.5). This could also contribute to the explanation of observed abundance of screw dislocations in experimental samples [4].

The force of screw dislocation interaction is $F(r_o) = -\frac{\partial E_T}{\partial r_o}$. This force can be determined as the function of $\frac{r_o}{r_c}$ from expression (13) and it is plotted in Figs. 4 and 5.

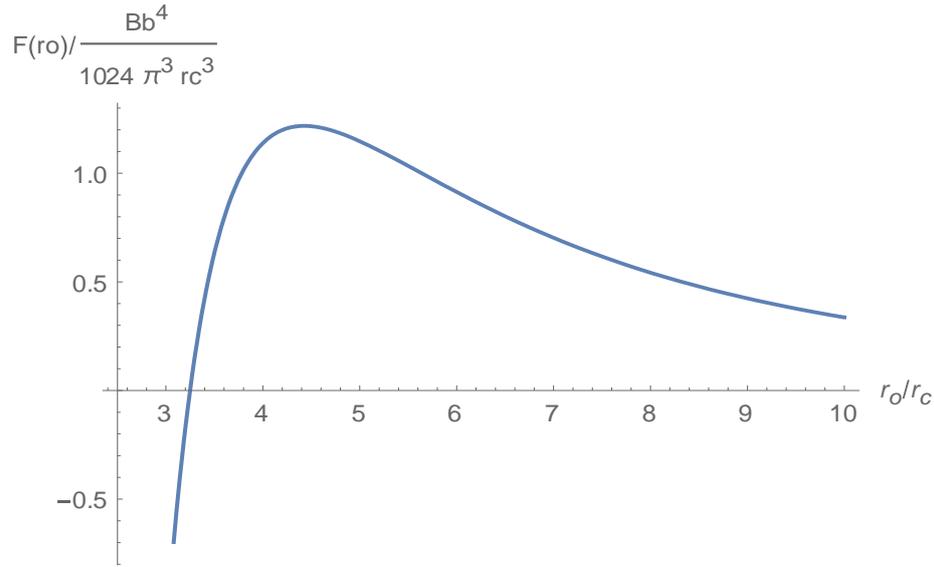

Fig. 4: The force of interaction between two screw dislocations of the same sign as the function of their separation $\frac{r_o}{r_c}$.

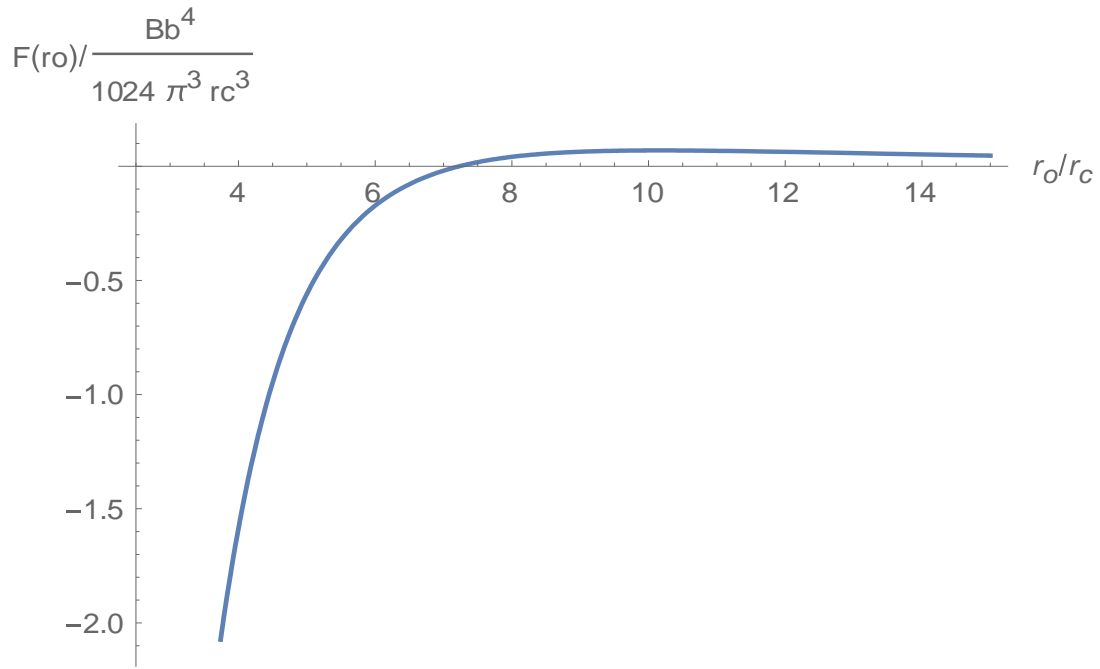

Fig. 5: The force of interaction between two screw dislocations of the opposite sign as the function of their separation $\frac{r_o}{r_c}$.

Now, let us compare force of edge dislocations interaction with the force of screw dislocation interaction. The force component $F_x$ of edge dislocations having Burgers vectors $b_1$ and $b_2$ can be written as [4]:

$$F_x = \frac{B b_1 b_2}{8\sqrt{\pi\lambda}} \frac{x_o}{|z_o|^{3/2}} exp\left(-\frac{x_o^2}{4\lambda|z_o|}\right), \tag{14}$$

with one dislocation situated at the coordinate origin and the second at position $(x_o, z_o)$. Parameter $\lambda$ in (14) is defined as $\lambda = \sqrt{\frac{K}{B}}$.

For $b_1 = b_2 = b$ at $x_o = z_o = b$, $\lambda \sim b$ the component $F_x$ given by (14) can be compared with the maximum of $F_{r_o}$ for dislocations of the same sign which is about $1.5\left(\frac{Bb^4}{1024\pi^4 r_c^3}\right)$. With $r_c \sim b$ we obtain:

$$\frac{F_x}{F_{r_o}} = \frac{1024\pi^3}{8\sqrt{\pi}} \frac{exp(-1/4)}{1.5} \sim 10^3. \tag{15}$$

Therefore we can conclude that the force between two screw dislocations is about three order smaller than the force between edge components. The force between screw dislocations evaluated at the anharmonic approximation can be taken as a correction only in cases when we deal with individual screw dislocations.

TGB phases modelled by screw dislocation walls are rather complex structure. Screw dislocation walls in TGB phases were treated e.g. in [14] as the continuous distribution of screw dislocations. The evaluation of their elastic energies in anharmonic approximation in [7] gave both their self-energies and interaction energies.

The nucleus of TGB phase in the form of filaments composed of finite blocks which are relatively rotated to each other by screw dislocation walls (see e.g. [16, 17]) is a more realistic case compared to TGB phases modelled just by blocks infinite in two dimensions [4, 12, 18] and separated by screw dislocation walls in the dimension parallel to the chiral axis. In filaments, the screw dislocation walls forming with edge dislocation walls finite dislocation loops. Edge dislocation walls mediate the nucleation of TGB phase in unperturbed smectic-A layers surrounding TGB filaments. When dealing with dislocation loops having both screw and edge dislocation segments, interactions between screw dislocations can be neglected with respect to the edge ones. In [16, 17] the energy of dislocation loops was taken principally as a line energy of loops, i.e. composed by the self-energies of edge and screw line dislocation segments, their interaction energies being neglected. The comparison of the elastic self-energy with the interaction energy of edge dislocations in [17] showed that the self-energy is about an order of magnitude greater than the interaction energy in the present study. The estimations also showed that interaction forces between screw dislocations are of the order smaller than interaction forces between edge dislocations. Thus we can conclude that our previous approximations in [16, 17] were valid.

### 5. Conclusions

Anharmonic correction to the elastic free energy of smectic A liquid crystal enables to evaluate the total elastic energy and the interaction force between two parallel screw dislocations as the function of their separation. The total elastic energy and the interaction force of parallel screw dislocations behaves differently as compared with solids. Anharmonic approximation leads to the existence of interaction terms proportional to second power of the product of Burgers vectors, i.e. to $(b^2 a_1^2 a_2^2)$.

Such terms do not depend on the sign of dislocation Burgers vectors and lead to the energy maximum changing the character of their interaction.

However, the interaction force between two screw dislocations is about three orders smaller than that of edge dislocations. One can conclude that the corrective interaction force between two screw dislocations calculated within an anharmonic approximation is important only in cases where we deal with long and isolated segments of screw dislocations in the smectic-A.

Funding

This work was supported by the Czech Ministry of Education, Youth and Sports [project number LTC19051 - MEYS] and Operational Programme Research, Development and Education financed by European Structural and Investment Funds [project No. SOLID21 - CZ.02.1.01/0.0/0.0/16_019/0000760].